\documentclass[a4paper,11pt]{article}
\usepackage{pos}
\usepackage{subcaption}
\usepackage{physics}
\usepackage[htt]{hyphenat}
\usepackage{soul} % \st : strikethrough

\usepackage{colortbl}
\definecolor{lightblue}{RGB}{202, 225, 255}
\definecolor{olivegreen}{RGB}{202, 255, 112}
\definecolor{darkolivegreen}{RGB}{85, 107, 47}
\definecolor{littledarkgreen}{RGB}{0,80,0}
\definecolor{firebrick}{RGB}{178,34,34}
\definecolor{darkslateblue}{RGB}{72,61,139}
\definecolor{midnightblue}{RGB}{25,25,112}
\definecolor{darkblue}{RGB}{0,0,139}
\definecolor{indigo}{RGB}{75,0,130}
\definecolor{dodgerblue}{RGB}{30,144,255}
\definecolor{mistyrose}{RGB}{255,228,225}
\definecolor{khaki}{RGB}{240,230,140}
\definecolor{royalblue}{RGB}{65,105,255}
\definecolor{mediumseagreen}{RGB}{60,179,113}
\definecolor{mediumspringgreen}{RGB}{60,179,113}
\definecolor{lime}{RGB}{0,255,0}
\definecolor{limegreen}{RGB}{50,205,50}
\definecolor{teal}{RGB}{0,128,128}
\definecolor{blueviolet}{RGB}{138,43,226}
\definecolor{lightmagenta}{RGB}{255,51,255}
\definecolor{background1}{RGB}{245,255,250}
\providecommand{\fm}{\;\mathrm{fm}}

\graphicspath{{./Figs/}}

\title{Performance of several Lanczos eigensolvers with \\ HISQ fermions}
\ShortTitle{Performance of several Lanczos eigensolvers with HISQ fermions}

\author*[a]{Hwancheol Jeong}
\author[b]{Carleton DeTar}
\author[a]{Steven Gottlieb}

\affiliation[a]{Department of Physics, Indiana University, Bloomington, Indiana 47405, USA}
\affiliation[b]{Department of Physics and Astronomy, University of Utah, Salt Lake City, Utah 84112, USA}

\emailAdd{sonchac@gmail.com}
\emailAdd{detar@physics.utah.edu}
\emailAdd{sg@indiana.edu}

\abstract{
  We investigate the state-of-the-art Lanczos eigensolvers available in the Grid and QUDA libraries. They include Implicitly Restarted Lanczos, Thick-Restart Lanczos, and Block Lanczos. We measure and analyze their performance for the Highly Improved Staggered Quark (HISQ) Dirac operator. We also discuss optimization of Chebyshev acceleration.
}

\FullConference{%
 The 38th International Symposium on Lattice Field Theory, LATTICE2021
  26th-30th July, 2021
  Zoom/Gather@Massachusetts Institute of Technology
}

\begin{document}
\maketitle

\section{Introduction}
\label{sec:intro}
Recent improvements in lattice QCD calculations have been achieved by making use of the eigenvalue spectrum of lattice operators, such as the lattice Dirac operator and the lattice Laplacian operator. The Lanczos iteration method is an eigensolver algorithm employed for that purpose \cite{Lanczos:1950zz}. It enables calculation of both eigenvalues and eigenvectors of a Hermitian matrix. It is most useful when a partial eigenvalue spectrum is required. Additional iterations provide more eigenmodes with higher accuracy.

There are several improvement methods for the Lanczos algorithm: (1) Polynomial acceleration applies a polynomial to the matrix of interest so that the Lanczos iteration converges faster with a transformed eigenvalue spectrum. Chebyshev polynomials are a popular choice~\cite{MR736453}. (2) There are two kinds of restarted Lanczos algorithms that take advantage of information learned from the previous run. The Implicitly Restarted Lanczos (IRL) algorithm suppresses unwanted eigenvalues at the restart \cite{Sorensen:1997aa}. Alternatively, the Thick-Restart Lanczos (TRL) algorithm suppresses or explicitly deflates converged wanted eigenvalues \cite{Wu_Simon:trl}. (3) There are approaches that utilize blocking to improve the computational efficiency of the Lanczos algorithm. The Block Lanczos (BL) algorithm can enhance the compute-to-communication ratio with the Split Grid method \cite{Jang:2019roq}.

Grid \cite{Boyle:2016lbp, grid:github} and QUDA \cite{Clark:2009wm, Babich:2011np, quda:github} are very popular and highly performing lattice QCD libraries, especially for modern parallel computing systems. For the Lanczos eigensolver, Grid has IRL and BL eigensolvers. QUDA has a TRL eigensolver and its block variant. All of them utilize Chebyshev polynomials to enhance the convergence.

We investigate these Lanczos eigensolver algorithms in the Grid and the QUDA libraries measuring and analyzing their performance for the Highly Improved Staggered Quark (HISQ) Dirac operator. In Section \ref{sec:simul}, we describe our simulation details. In Section \ref{sec:lanczos}, we describe some basics of the Lanczos iteration method. Section \ref{sec:poly} describes the general strategy of utilizing Chebyshev polynomials and studies its optimization. In Sections \ref{sec:irl} and \ref{sec:trl}, we illustrate ideas of IRL and TRL, respectively, and discuss their optimizations. In Section \ref{sec:bl}, we examine the performance of BL with the Split-Grid method. We conclude in Section \ref{sec:conc}.

\section{Simulation details}
\label{sec:simul}
We calculate eigenvalues and eigenvectors of the low modes of the massless HISQ Dirac operator $D$. The Lanczos eigensolvers are run using a Hermitian variation $D^\dagger D$, which is (semi)-positive definite that allows even-odd splitting. The eigensolver performance is measured in elapsed time taken by the eigensolver routine itself. We use an $N_f=2+1+1$ MILC HISQ gauge configuration, with lattice size  $24^3 \times 64$, lattice spacing $a \approx 0.12 \fm$, and $am_l / am_s = 0.00507 / 0.0507$ \cite{Bazavov:2012xda}.

We run Grid on Intel CPUs and QUDA on NVIDIA GPUs, where they perform their best so that we can focus on algorithmic performance. All Grid runs are done on a single node of dual Intel Xeon E5-2650v2 system that has $2 \times 8 = 16$ CPU cores, where we use only half, {\it i.e.,} eight, of them for performance stability. We apply OpenMP for their parallelization, except for the BL eigensolver in Section \ref{sec:bl}, where we apply MPI. QUDA runs are done on the same system but with two NVIDIA K80 GPUs.

\section{Lanczos algorithm and Lanczos iteration method}
\label{sec:lanczos}
Let us consider an $m \times m$ Hermitian matrix $A$, and an $m$-dimensional vector $b$. The Gram-Schmidt process on a set of vectors $\{ b, Ab, A^2 b, \cdots, A^{n-1} b \}$ for $n \leq m$, which spans a Krylov subspace $\mathcal{K}_n (A,b)$, gives a three term recurrence relation
\begin{equation}
  \label{eq:lanczos_recurr}
  t_{i+1,i}\, q_{i+1} = A q_i - t_{i,i}\, q_i - t_{i-1,i}\, q_{i-1} \,,
\end{equation}
where $q_i$ are orthonormal basis vectors of $\mathcal{K}_n (A,b)$, and $t_{ij} = \braket{q_{i\,}}{\,A q_j}$. Combining recurrences of $i=1,\cdots,n$, we have
\begin{equation}
  \label{eq:lanczos}
  T_n = Q_n^\dagger A Q_n \,,
\end{equation}
where $Q_n = (q_1|q_2|\cdots|q_n)$, and $T_n = (t_{ij})$ is an $n \times n$ Hermitian tridiagonal matrix.

The Lanczos algorithm is to construct $T_n$ and $Q_n$ by iterating the Lanczos recurrence relation Eq.~\eqref{eq:lanczos_recurr} \cite{Lanczos:1950zz}. When $n = m$, $Q_m$ is a unitary transformation, so $T_m$ has the same eigenvalue spectrum as $A$. Even for $n < m$, eigenvalues of $T_n$ are approximate to some $n$ eigenvalues of $A$ \cite{Trefethen}. They converge to true eigenvalues of $A$ as $n$ increases. The largest, the smallest, or the most isolated (the least dense) extreme eigenvalues appear first and converge first. When $k (< m)$ eigenvalues are wanted, we need to perform $n$ ($k \leq n \leq m$) iterations of the Lanczos recurrence (Eq.~\eqref{eq:lanczos_recurr}) with the result that $T_n$ has eigenvalues approximate to $k$ eigenvalues of $A$ of interest within the target precision. 

A QR iteration is usually employed to diagonalize a tridiagonal matrix $T_n$ \cite{Wilkinson:1968aa, Wang:2002aa}. Then its eigenvalue decomposition gives $T_n = V_n \Lambda V_n^\dagger$, where $\Lambda$ is a diagonal matrix with eigenvalues $\lambda_i$ of $T_n$, and $V_n$ is composed of their corresponding eigenvectors $v_i$. With this, we calculate eigenvector estimates (called Ritz vectors) $w_i$ and eigenvalue estimates (called Ritz values) $\tilde{\lambda}_i$ by
\begin{equation}
  \label{eq:ritz_vector_value}
  w_i = Q_n v_i \,, \quad \tilde{\lambda}_i = \frac{\mel{w_i}{A}{w_i}}{\braket{w_i}{w_i}} \,.
\end{equation}
The convergence of each eigenvector is measured by a residual
\begin{equation}
  \label{eq:lanczos_residual}
  \frac{ \norm{ A\!\ket{w_i} - \tilde{\lambda}_i\!\ket{w_i} } }{\sqrt{\braket{w_i}{w_i}} } \,,
\end{equation}
or dividing it by some normalization factors. Note that $\tilde{\lambda_i}$ can be different from $\lambda_i$ since we usually apply a polynomial to $A$, as will be discussed in Section \ref{sec:poly}.

\section{Chebyshev acceleration}
\label{sec:poly}
The convergence of a Lanczos iteration depends on the eigenvalue density. Less dense (including the largest and the smallest) eigenvalues appear early and converge fast. One can control the density of eigenvalues, and, thus, the convergence of Lanczos iteration, by applying a polynomial $\mathcal{P}$ to the matrix $A$. It maps eigenvalues $\lambda_i$ of $A$ into $\mathcal{P}(\lambda_i)$ but preserves their eigenvectors.

Chebyshev polynomials $C_p(x)$ of degree $p$ are bounded within $[-1,1]$ for $|x| \leq 1$, while diverging rapidly for a high $p$ as $|x|$ increases for $|x| > 1$ \cite{wiki:chebyshevPoly}. We can transform it to map unwanted eigenvalues into the bounded region and wanted eigenvalues into the diverging region. This Chebyshev acceleration can boost the convergence of Lanczos iteration. With suitable choices of the location of the bounded region and the polynomial degree, its benefit more than compensates for the cost of applying the polynomial.

Eigenvalues of $D^\dagger D$ are non-negative, and our usual interests lie in its low modes. We transform the domain of $C_p(x)$ so that $[-1,1]$ becomes $[\alpha, \beta]$ that covers the unwanted eigenvalues. The value of $\alpha$ should be set to be greater than the largest desired eigenvalue, and $\beta$ to be greater than the largest eigenvalue in the whole spectrum. A few power iterations can help to estimate the size of the largest eigenvalue. Choosing $\alpha$ needs some heuristics, but we may use the same value within a gauge ensemble.

\begin{figure}[tb]
  \captionsetup[subfigure]{aboveskip=0.2em,belowskip=0.5em}
  \captionsetup{aboveskip=-0em,belowskip=0em}
  \centering
  \begin{subfigure}[t]{.32\linewidth}
    \includegraphics[width=\linewidth]{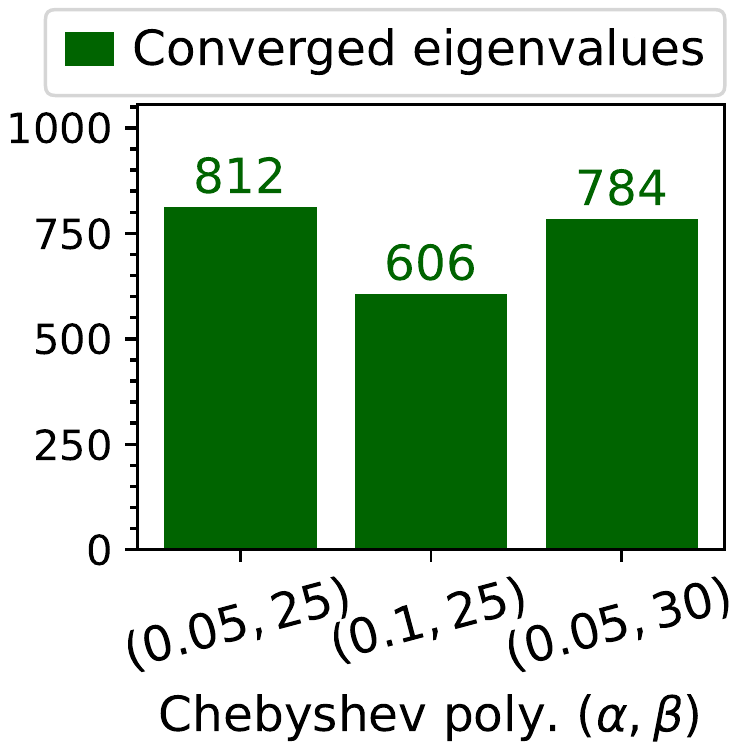}
    \caption{}
    \label{fig:poly_ab}
  \end{subfigure}
  \begin{subfigure}[t]{.64\linewidth}
    \includegraphics[width=\linewidth]{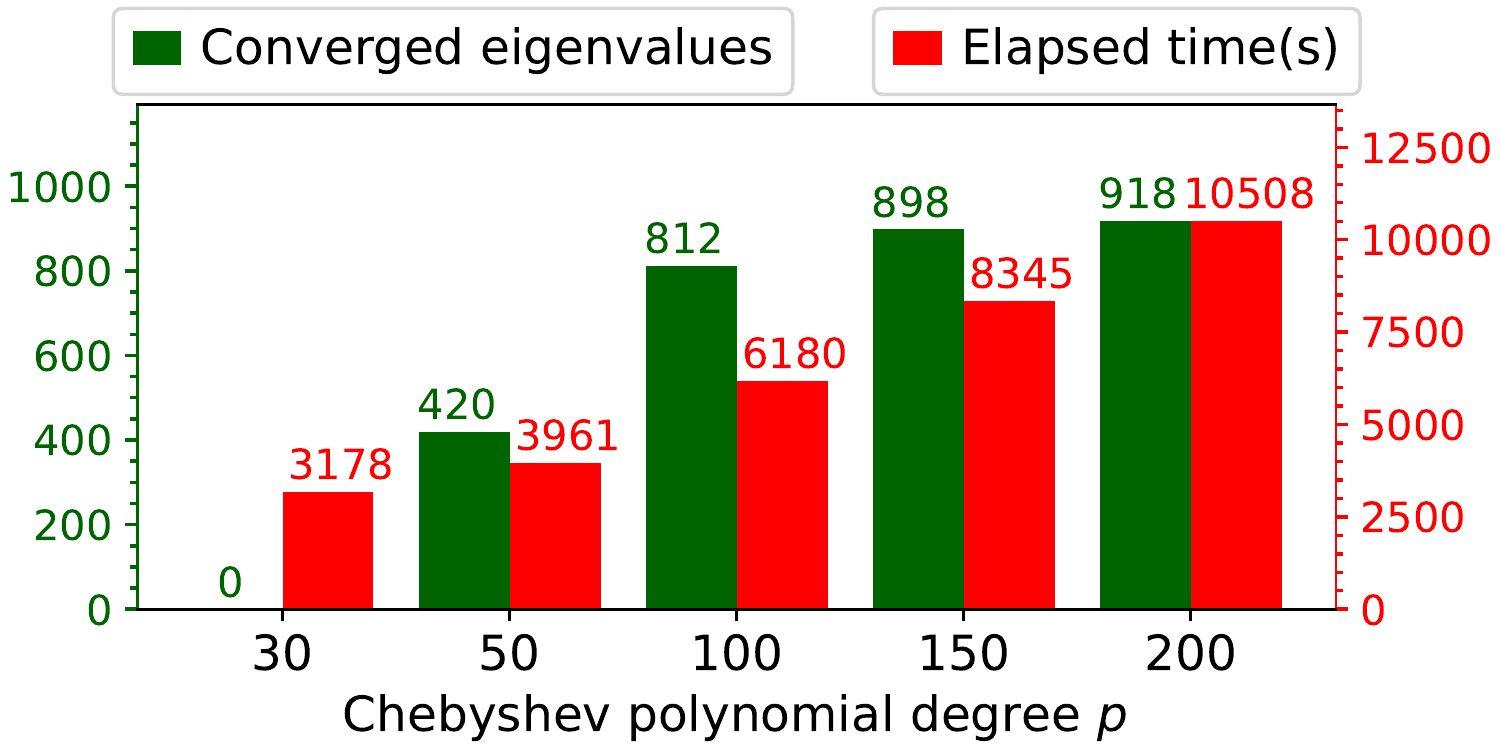}
    \caption{}
    \label{fig:poly_n}
  \end{subfigure}
  \caption{The convergence of Chebyshev accelerated Lanczos iterations for different Chebyshev parameters $(\alpha, \beta)$ (left) and degree $p$ (right).}
  \label{fig:poly}
\end{figure}
In Fig.~\ref{fig:poly}, we show the convergence of Lanczos iterations with different Chebyshev acceleration parameters. Here, Grid's \texttt{ImplictlyRestartedLanczos} eigensolver is used with a little modification not to restart. We run 1000 Lanczos iterations, and the convergences are checked by Eq.~\eqref{eq:lanczos_residual} in double precision. Figure~\ref{fig:poly_ab} presents the number of converged eigenvalues for different $\alpha$ and $\beta$ values. The same Chebyshev polynomial degree $p = 100$ is used here, which means all the runs take the same time in principle. On this gauge configuration, the largest eigenvalue of $D^\dagger D$ is around 21, and the 1000th eigenvalue is around 0.046. The result indicates that we get the best convergence when both $\alpha$ and $\beta$ are the closest to the 1000th eigenvalue and the largest eigenvalue, respectively. The 800th eigenvalue is around 0.034. Hence, $\alpha = 0.04$ might give a better convergence on this gauge configuration. However, a tight choice may not work for other gauge configurations in the same ensemble.

In Fig.~\ref{fig:poly_n}, we present the results for different polynomial degrees $p$. Here, we used $(\alpha, \beta)$ = $(0.05, 25)$, the best pair found in Fig.~\ref{fig:poly_ab}. Although higher $p$ makes more eigenvalues converge, it demands more computations. In this example, $p=100$ gives the best efficiency. In general, the most efficient $p$ depends on the number of eigenvalues we want. We need heuristics to find an optimized $p$.

Every improved Lanczos algorithm discussed in this paper utilizes Chebyshev acceleration. We use $(\alpha, \beta) = (0.05, 25)$, and $p = 50 \text{ or } 100$.

\section{Implicitly Restarted Lanczos}
\label{sec:irl}
A basic Lanczos algorithm must carry on its iteration by increasing the Krylov subspace dimension $n$ until eigenvalues of the desired number converge. As $n$ increases, it takes more memory to store the eigenvectors as well as more computing time for the iteration\footnote{In practice, Lanczos vectors $q_i$'s orthogonality gets inexact as iteration continues due to the accumulation of numerical errors. One can correct it by re-orthogonalizing all $q_i$'s per some period of iterations. This is expensive for a large $n$.} and for the diagonalization of a larger $T_n$. Moreover, it is difficult to anticipate an optimized (or the smallest) $n$ without monitoring the convergence, which includes diagonalizing $T_i$ ($i<n$) and computing eigenvectors and their residuals.

Suppose we have performed $n$ Lanczos iterations for Hermitian matrix $A (= C_p(D^\dagger D))$ with a starting vector $b$. The Implicitly Restarted Lanczos (IRL) algorithm allows us to restart from the $(n-k)$-th step of the Lanczos algorithm for $A$ with a new starting vector $\tilde{b} = (A-\mu_k) \cdots (A-\mu_1) b$, where the $\{\mu_i, i=1,\ldots k\}$ are some eigenvalues \cite{Sorensen:1997aa}. The Krylov subspace $\mathcal{K}_n( A, \tilde{b} )$ for the restarted run does not contain eigenvectors corresponding to the eigenvalues ${\mu_i}$. It not only excludes unwanted eigenvalues and eigenvalues near them, but it also improves the convergence of the remaining wanted eigenvalues far from the $\{\mu_i\}$.

When we are interested in only the $t$ lowest eigenmodes, a typical implementation of IRL runs the Lanczos iteration to the $n$-th step, does the implicit restart that excludes the $k$ largest eigenvalues found in the previous run, and repeats them until the $t$ smallest eigenvalues converge. In this way, we can constrain the Krylov subspace dimension to any $n > t$, as much as the system memory allows. However, since $k$ is usually set to make $n-k$ slightly larger than $t$, $n$ determines how many restarts we need to do. Each restart must perform the QR decomposition $k$ times and do the basis rotations using them. This cost can be comparable to that of many Lanczos iterations. Hence, we need to find an optimized $n$ that maximizes the restarting efficiency.

\begin{figure}[tb]
  \captionsetup[subfigure]{aboveskip=0.2em,belowskip=0.5em}
  \captionsetup{aboveskip=-0em,belowskip=0em}
  \centering
  \begin{subfigure}[t]{.95\linewidth}
    \includegraphics[width=\linewidth]{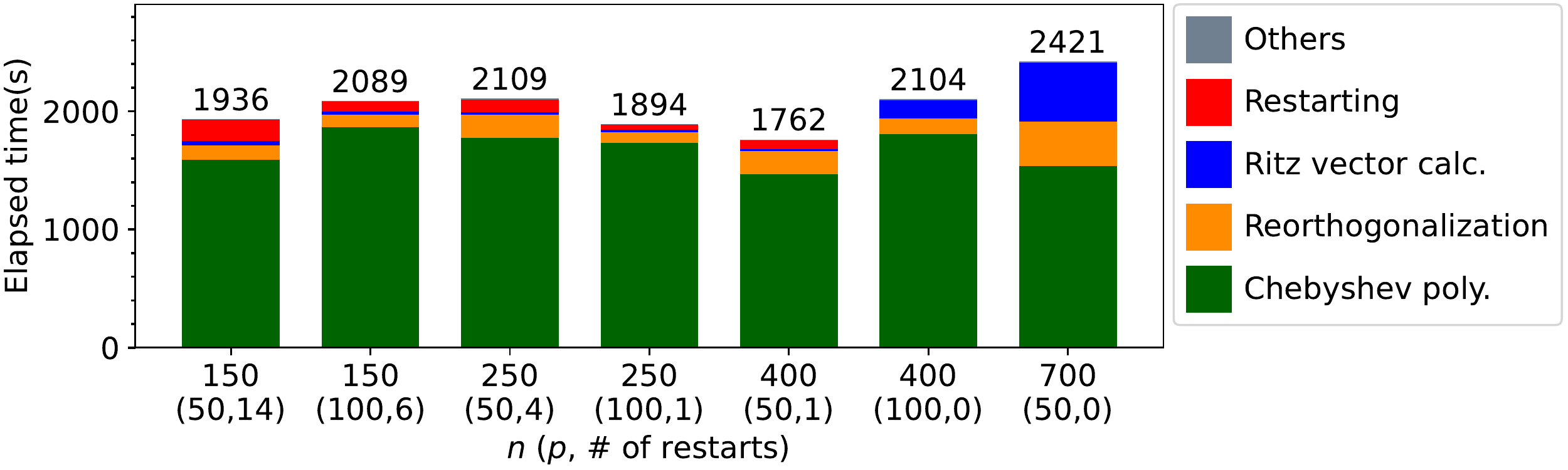}
    \caption{(Number of eigenvalues) = 100}
    \label{fig:irl_100}
  \end{subfigure}
  \begin{subfigure}[t]{.95\linewidth}
    \includegraphics[width=\linewidth]{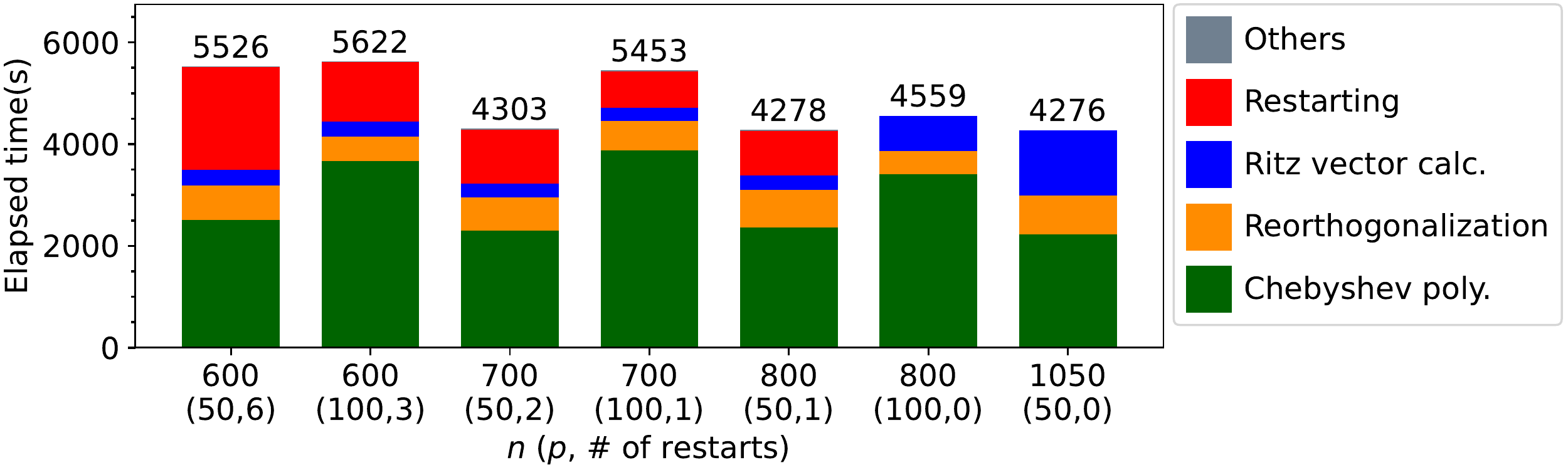}
    \caption{(Number of eigenvalues) = 500}
    \label{fig:irl_500}
  \end{subfigure}
  \caption{Performance of the Grid IRL eigensolver for various Krylov subspace dimensions $n$ with Chebyshev polynomial degrees $p=50, 100$.}
  \label{fig:irl}
\end{figure}
The Grid library has an IRL eigensolver named \texttt{ImplicitlyRestartedLanczos}. In Fig.~\ref{fig:irl}, we present its performance profiles for various Krylov subspace dimensions $n$ with Chebyshev acceleration of $p=50,100$. For non-restarted cases, a modified version mentioned in Section \ref{sec:poly} is used. Note that in the Lanczos iteration, eigenvalues converge simultaneously, not one by one in order. Hence, it could be less efficient for calculating a small number of eigenvalues. In Fig.~\ref{fig:irl_100}, where we calculate $100$ eigenvalues, the non-restarted Lanczos converges at $n = 400$ and $n = 700$ for $p = 50$ and $p = 100$, respectively. In a (relatively) slowly converging case like this, IRL can perform better than the basic non-restarted Lanczos. For both $p=50$ and $100$, we obtain the best performance by restarting once.

On the other hand, in Fig.~\ref{fig:irl_500}, where we calculate $500$ eigenvalues, the non-restarted Lanczos takes the least time. Here, $n=800,1050$ are only about twice $500$, so the restarting costs for $n=600,700,800$ are comparable to or more expensive than running Lanczos on a bigger Krylov subspace. However, the differences between $n=700,800,1050$ are not huge for $p=50$, so one may choose $n=700$ to save memory. In addition, unlike the non-restarted Lanczos where the convergence is not guaranteed for a given $n$, IRL assures its convergence when the same $n$ is used for other gauge configurations, even though it may need more restarts.

One last remark is that we should optimize the Chebyshev acceleration parameters together with $n$. In Fig.~\ref{fig:irl_100}, the non-restarted Lanczos converges faster with $p=100$ than $p=50$, while the IRL's best performance comes with $p=50$.

\section{Thick-Restart Lanczos}
\label{sec:trl}
In this section, we investigate another restarted Lanczos algorithm called Thick-Restart Lanczos (TRL) \cite{Wu_Simon:trl}. Unlike IRL, which suppresses unwanted eigenvectors to improve convergence, TRL suppresses (nearly) converged wanted eigenvectors from the Krylov subspace for the restarted run.

One can rewrite Eq.~\eqref{eq:lanczos} as
\begin{equation}
  \label{eq:lanczos_diag}
  \Lambda_k = ( Q_n V_k )^\dagger A\, Q_n V_k = {W_k}^\dagger A\, W_k \,,
\end{equation}
where $V_k$ is composed of some $k(<n)$ eigenvectors of $T_n$, and $W_k \equiv (w_1 | w_2 | \cdots | w_k)$ is composed of the corresponding $k$ eigenvector estimates (Ritz vectors) of $A$. $\Lambda_k$ is a diagonal matrix of $k$ eigenvalues of $T_n$ that are eigenvalue estimates (Ritz values) of $A$. Note that $w_i$'s are basis vectors of the same Krylov subspace $\mathcal{K}_n (A,b)$. Hence, we can replace $Q_n$ with $W_k$, and $T_n$ with $\Lambda_k$ in Eq.~\eqref{eq:lanczos}.

The TRL algorithm restarts the Lanczos iteration upon $W_k$ and $\Lambda_k$, but by appending the vector $q_{n+1}$.\footnote{One cannot start the Lanczos algorithm with an eigenvector, because it generates a rank-$1$ Krylov subspace.} One problem is that $w_i$'s are not orthogonal to $q_{n+1}$, so $\braket{A q_{n+1}}{w_i}$ terms survive in the Gram-Schmidt process generating $q_{n+2}$. This results in a arrow-like matrix $T_{n+1}$. However, thanks to the symmetry of $A$, the generation of $q_{n+2+i}$ $(i \geq 1)$ returns to the normal three-term Lanczos recurrence. Hence, we need to deal only with diagonalizing the arrow-like matrix, which can be done in several ways including the general matrix diagonalization.

Now that $q_{n+2}$ is orthogonal to $w_1, \cdots, w_k$, approximate eigenvectors of $A$, the restarted run searches on the Krylov subspace orthogonal to them. It focuses on finding other eigenvectors rather than $w_i$'s, though it still slowly improves $w_i$'s convergences as well. With a proper choice of $k$, it can balance the convergence of eigenvalues efficiently.

The QUDA library has a TRL eigensolver named \texttt{TRLM}. This eigensolver determines $k$ as the number of converged eigenvalues in the desired precision plus the half of the remaining eigenvalues that are still converging. It also implements the locking method, which explicitly deflates some converged eigenvectors from the restarted run. \texttt{TRLM} locks eigenvectors converged to the machine precision. It reduces memory usage and computing cost.

\begin{figure}[tb]
  \captionsetup[subfigure]{aboveskip=0.2em,belowskip=0.5em}
  \captionsetup{aboveskip=-0em,belowskip=0em}
  \centering
  \begin{subfigure}[t]{.95\linewidth}
    \includegraphics[width=\linewidth]{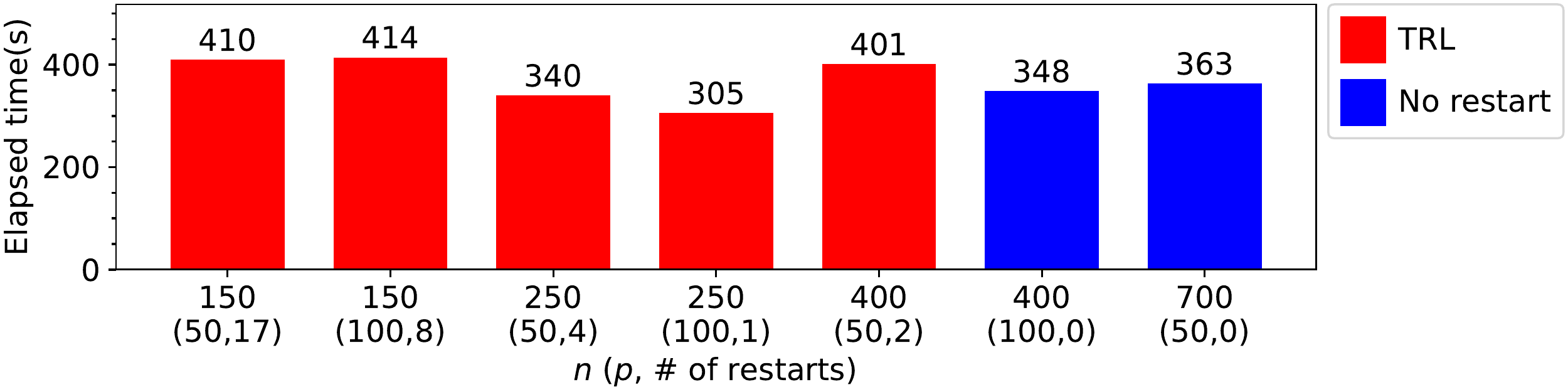}
    \caption{(Number of eigenvalues) = 100}
    \label{fig:trl_100}
  \end{subfigure}
  \begin{subfigure}[t]{.95\linewidth}
    \includegraphics[width=\linewidth]{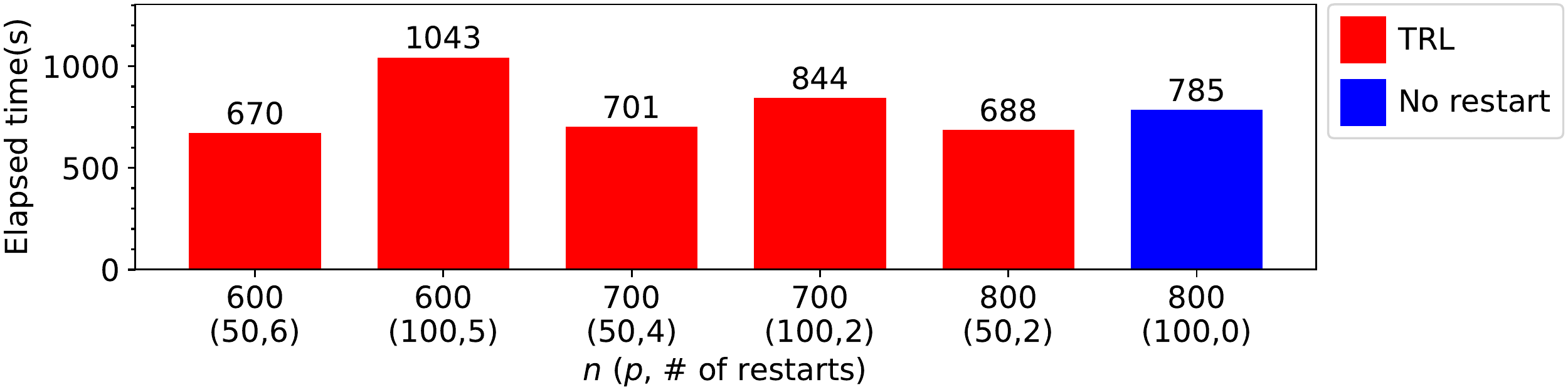}
    \caption{(Number of eigenvalues) = 500}
    \label{fig:trl_500}
  \end{subfigure}
  \caption{Performance of the QUDA TRL eigensolver for various Krylov subspace dimensions $n$ with Chebyshev polynomial degrees $p=50, 100$.}
  \label{fig:trl}
\end{figure}
In Fig.~\ref{fig:trl}, we present QUDA \texttt{TRLM}'s performance for various Krylov subspace dimensions $n$ with Chebyshev acceleration of $p=50,100$. In both Fig.~\ref{fig:trl_100} and Fig.~\ref{fig:trl_500}, we find TRL can perform better than the non-restarted Lanczos with optimization. One benefit of TRL, compared with IRL, is that its restarting cost can be absorbed into the Ritz vector calculation. That might be why TRL takes the least time for $n=600$ with six restarts in Fig.~\ref{fig:trl_500}.

Restarted Lanczos algorithms (both TRL and IRL) allow us to use less memory. It is especially useful for GPU systems, where memory is usually lacking. Figure~\ref{fig:trl_500} does not include the non-restarted Lanczos result for $p=50$ because it does not converge within the given GPU memory size.

\section{Block Lanczos with Split Grid}
\label{sec:bl}
The Block Lanczos (BL) algorithm runs the Lanczos iteration with multiple ($u$) starting vectors $\{ b_1, b_2, \cdots, b_u \}$, where $b_i$'s are orthogonal to each other \cite{Jang:2019roq}. At each $j$-th iteration of BL, a block version of the Lanczos recurrence Eq.~\eqref{eq:lanczos_recurr} constructs $u$ basis vectors $q^1_j, \cdots, q^u_j$ of the combined Krylov subspace $\mathcal{K}_n (A,b_1) \cup \cdots \cup \mathcal{K}_n (A,b_u)$ by orthogonalizing $A q^1_{j-1}, \cdots, A q^u_{j-1}$ simultaneously. After $n$ iterations, it transforms $A$ into a block-tridiagonal matrix $T_n$ of dimension $(nu \times nu)$.

The convergence of a BL iteration is usually slower than that of the basic Lanczos for the same subspace dimension $\tilde{n} = nu$, because the Krylov subspace dimension $n$ per starting vector $b_i$ is smaller than $\tilde{n}$. However, on a parallel computing system, BL may outperform the basic Lanczos by parallelizing the multiplications $A q^i_{j-1}$ for $i = 1, \cdots, u$. The Split Grid method is one way of doing it \cite{Jang:2019roq}. It splits a communication grid such as MPI and distributes parallel jobs into the split grids, so that each job runs with a lower surface-to-volume ratio.

\begin{figure}[tb]
  \captionsetup[subfigure]{aboveskip=0.2em,belowskip=0.5em}
  \captionsetup{aboveskip=-0em,belowskip=0em}
  \centering
  \begin{subfigure}[t]{.95\linewidth}
    \includegraphics[width=\linewidth]{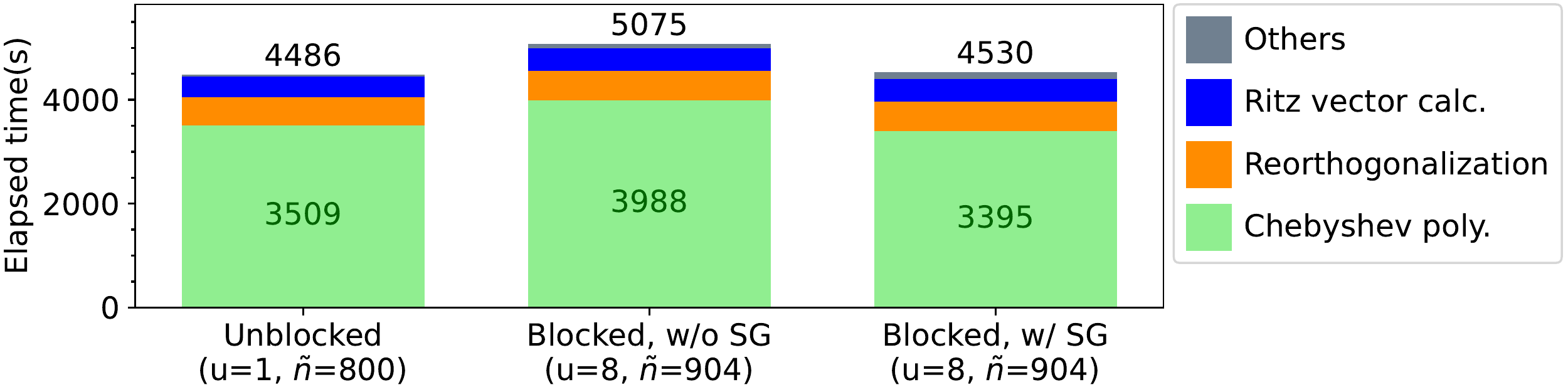}
    \caption{(Number of eigenvalues) = 500}
    \label{fig:bl_500}
  \end{subfigure}
  \caption{Performance of the Grid BL eigensolver with the Split Grid method (SG), compared with unblocked basic Lanczos and BL without the Split Grid method.}
  \label{fig:bl}
\end{figure}
The Grid library has a BL eigensolver implementing the Split Grid method, named \texttt{ImplicitlyRestartedBlockLanczos}.\footnote{Although it provides both non-restarted and implicitly restarted versions of BL, the latter is not pursued.} In Fig.~\ref{fig:bl}, we compare its performance with the unblocked basic Lanczos and BL without the Split Grid method. The block size for BL is set to $u = 8$. In this measurement, we run $8$ MPI processes connected by an intra-node network. We calculate $500$ eigenvalues, for which the unblocked Lanczos converges at $\tilde{n} = 800$, while the BL converges at $\tilde{n} = 904$. Hence, without the Split-Grid method, the BL algorithm itself is slower than the basic Lanczos algorithm, as the results show.

However, the Split Grid method can enhance the BL's performance. In Fig.~\ref{fig:bl}, BL's performance is improved by 12\% with the Split Grid method, even on the intra-node network. Although it is still a little slower than the unblocked Lanczos, its advantage could become significant on a slow inter-node network.\footnote{It is under investigation. A preliminary result shows that with $8$ MPI inter-nodes, BL with Split Grid gives around $200\%$ better performance than the unblocked Lanczos.}

\section{Conclusion}
\label{sec:conc}
We have discussed improved Lanczos algorithms and their optimizations. A well-tuned Chebyshev polynomial improves the Lanczos iteration's convergence significantly. All other improved Lanczos algorithms utilize it. Restarted Lanczos algorithms, such as Implicitly Restarted Lanczos and Thick-Restart Lanczos, do not always perform better than the non-restarted Lanczos algorithm. Still, we can optimize them to perform better than or comparable to non-restarted Lanczos. It is advantageous for small memory systems such as GPUs. The performance of Block Lanczos utilizing the Split Grid method is comparable to that of the unblocked Lanczos on an intra-node network. It may perform better on inter-node networks.

% Acknowledgment
\acknowledgments{
  We would like to thank D.~Howarth, Y.~Jang, and C.~Jung for useful discussions. This research was supported by the Exascale Computing Project (17-SC-20-SC), a collaborative effort of the U.S. Department of Energy Office of Science and the National Nuclear Security Administration. We also gratefully acknowledge support from the Department of Energy grant DE-SC0010120 and from the National Science Foundation grant PHY20-13064. Finally, we thank the developers who support Grid and QUDA whose names can be found at the respective websites for the software.
}

\bibliographystyle{JHEP-hc}
\bibliography{ref}

\end{document}